\documentstyle[aps,prl,preprint]{revtex}
\begin{document}
\draft
\title{Large-Scale Numerical Evidence for Bose\\
Condensation in the $S=1$\\
Antiferromagnetic Chain in a Strong Field}
\author{Erik S.\ S\o rensen$^a$ and Ian Affleck$^{a,b}$ }
\address{$^{(a)}$Department of Physics and $^{(b)}$Canadian Institute
for Advanced Research}
\address{
University of British Columbia, Vancouver, BC, V6T 1Z1, Canada}
\date{\today\, \, \small UBCTP-93-008}
\maketitle
\begin{abstract}
Using the recently proposed density matrix renormalization group
technique~\cite{white1} we show that the magnons in the $S=1$
antiferromagnetic Heisenberg
chain effectively behaves as bosons that condense
at a critical field $h_c$. We determine the
spin-wave velocity, $v=2.49(1)$, as well as the gap
$\Delta=0.4107(1)J$.
\end{abstract}
%====================================================================
% BODY OF PAPER

It is by now well established both experimentally~\cite{expgap} and
theoretically~\cite{numgap,tak1,white2,takahashi,jphysc}
that the $S=1$ antiferromagnetic chain has a gap, $\Delta$, to a
triplet excitation above a singlet ground-state.
Thus the magnetization, $M$,
remains strictly zero up to a critical field $h_c=\Delta$.  For the
quasi-one-dimensional system NENP
the critical field is found to be about
10T~\cite{fgap}
for a field applied along the symmetry axis.
The behavior of $M(h)$ just
above $h_c$ has been the subject of some theoretical
work~\cite{tsvelik,nomura,ian2}.  The existing
experimental results~\cite{katsumata}
appear to be dominated by extraneous effects such as
an off-diagonal alternating component
of the gyromagnetic tensor~\cite{gyro},
inter-chain coupling, anisotropies and impurities.

This problem can be solved
using a boson quasi-particle model involving
triplet magnons with repulsive
interactions for parallel spins~\cite{ian2}.
This model
predicts that $h_c=\Delta$.
(We set the Bohr magneton and $g$-factors to
one.) At this field the ``rest-mass energy''
of the magnons is exactly
cancelled by their Zeeman
energy and one-dimensional Bose condensation
occurs.  What prevents a
catastrophe from occurring at $h_c$ is the inter-magnon
repulsion which governs
the behavior of $M(h)$ above $h_c$.  To calculate
$M(h)$ we need to calculate
the energy, $E(M)$  of a very dilute system of $M$
polarized magnons in their ground-state.
$M(h)$ is then found by inverting the
equation: $h= dE/dM$.  It was argued in Ref. \onlinecite{ian2}
that the behavior of $E(M)$
is the same as for a system of
non-relativistic non-interacting {\it fermions}:
\begin{equation}
E=(\Delta-h)M+L\int_{-k_F}^{k_F}{dk \over 2\pi}{v^2k^2\over
2\Delta}.
\label{eq:1}
\end{equation}
Here $v$ is the velocity, determined from the single
magnon dispersion relation at low energies:
\begin{equation}
\omega (k) =
\Delta + v^2(k-\pi )^2/2\Delta + O([k-\pi ]^3),
\end{equation}
$L$ is the
length of the system and $k_F$ is determined from the particle number:
\begin{equation}
M=L\int_{-k_F}^{k_F}{dk \over 2\pi}=Lk_F/\pi .
\end{equation}
This gives, $E=M(\Delta-h)+(v\pi )^2M^3/6\Delta L^2+O(M^4)$ and hence
\begin{equation}
M/L=\sqrt{(h-\Delta )2\Delta}/\pi v,
\label{eq:ml}
\end{equation}
up to terms of higher order in $h-\Delta$.

This formula was
first given based on a non-interacting fermion model of
magnons~\cite{tsvelik}.
It was later argued to also arise from an {\it interacting boson}
model~\cite{ian2}
sufficiently close to $h_c$, and hence to be exact.  It is expected to
be valid for very general
short-range repulsive interactions between the
spin-polarized magnons.
It should hold for arbitrarily weak interactions for long enough
chains.
The reason is that
when the average inter-magnon
spacing is very large compared to the range of
the repulsive interaction the multi-magnon wave-function,
$\Psi_M(x_1,x_2,...x_M)$,
can be approximated by a free fermion (Bloch)
wave-function multiplied
by the sign function $\epsilon (x_1,x_2,...x_M)$
which has the value $\pm 1$
and changes sign whenever two particles are
interchanged of the following form
\begin{equation}
\Psi_M(x_1,x_2,...x_M)=\frac{1}{\sqrt{M!}}\epsilon (x_1,x_2,...x_M)
\sum_{P(i_1\ldots i_M)}\prod_{i=1}^M
\psi_1(x_{i_1})\psi_2(x_{i_2})\ldots\psi_M(x_{i_M}){\rm sgn}P.
\label{eq:wf}
\end{equation}
Here $P$ denotes the permutation and ${\rm sgn}P$ the sign
of the permutation. $\psi_i(x)$ is a single-particle non-interacting
wave-function depending on the wave-vector $k_i$.
$\Psi_M$ is symmetric as required by Bose
statistics, is a solution of the non-interacting
Schroedinger equation almost
everywhere
(ie. except where two or more particles intersect) and vanishes
whenever two or more particles come together.
As such it is expected to
become exact in the dilute
limit and hence to give exactly the magnetization
as $h \to h_c$.

Until very recently it has been essentially
impossible to test this prediction by
numerical simulations.  The reason is that extremely long systems are
required.
To simulate the $M$-magnon problem in the dilute regime we need a
length $L>>M \xi$ where $\xi \approx 6$
is the correlation length or the
approximate range
of the inter-magnon interaction.  In practice it appears that
$L \geq 30 M$ is required.
Before last year,  the longest chains that had
been studied accurately had $L=32$
so good results were only available for a
single magnon and it was impossible
to study magnon interactions.  This
situation has now completely
changed thanks to a breakthrough in the real-space
renormalization group technique made by Steven White~\cite{white1}
which makes it quite
feasible to study chains of length 100 or longer using a
density matrix renormalization group (DMRG) approach.
We will present results here
on chains of length up to 100 containing up to three magnons.
White and Huse~\cite{white2} obtained related numerical results
independently. We analyze our results in a different way which
establishes Eq.~(\ref{eq:ml}).
Our results
indicate quite convincingly
that the lowest energy two or three magnon state
has the form discussed above; namely the lowest-energy free fermion
wave-function multiplied by the sign function.
We establish this result in
two ways. Firstly we study $S^z(x)=\sum_{j=1}^M\delta(x-\hat x_j)$,
where $\hat x_j$ is the position operator for the j'th particle,
showing that it has
the expected form:
\begin{equation}
<S^z (x)> =M\int
dx_2dx_3...dx_M|\Psi_M(x,x_2,x_3,...x_M)|^2.
\label{eq:4}
\end{equation}
Secondly, we study the
finite-size dependence
of the energy of the multi-magnon ground-state, showing
that it behaves as:
\begin{equation}
E(M) \approx \sum_{i=1}^M\omega (k_i).
\label{eq:5}
\end{equation}
Holding $M$ fixed and taking $L \to \infty$, the
$k_i$'s are $O(1/L)$ so the above formula gives
$E(M,L)=M(\Delta-h)+a(M)/L^2$.
The corrections to Eq. (\ref{eq:5}) from
inter-magnon interactions
are expected to be $O(L^{-3})$.  This power of $L$
can be most easily understood in the $M=2$ case.
The wave-function contains a
$L^{-1}$ normalization factor
but is $O(L^{-2})$ over the entire region where
the interaction is non-negligible since it vanishes proportional to
$k(x_1-x_2)$ with $k$ of $O(L^{-1})$.
The power $L^{-3}$ arises from squaring
the wave-function and picking up a factor
of $L$ from the integral over the
center-of-mass position.
We verify that the corrections to Eq. (\ref{eq:5}) are
indeed of this order.

The DMRG
method {\it does} create certain technical problems because
of the fact that
implementing it optimally requires studying chains with
open,
rather than periodic boundary conditions.
Such chains have $S=1/2$ excitations
localized near the ends which have
been the subject of a number of experimental
and theoretical studies~\cite{endex}.
For our purposes they are just a minor annoyance.
They are also found to have repulsive interactions
with the bulk magnons
for parallel spins.  Thus
the magnon wave-functions essentially
obey vanishing boundary conditions at the
ends, with additional corrections of $O(L^{-3})$ to the energy,
Eq. (\ref{eq:5}), from the
magnon-end excitation interactions. We also considered more general
boundary conditions for the magnon wave-functions. These were found to
change the energy only to $O(L^{-3})$ since they lead to negligible
changes in the wave-function.

The DMRG method for open chains leaves only two good quantum numbers
the total $S^z$ component, $S^z_T$, and the parity, $P$. These are
conserved under iteration
and it is therefore possible to work within a
subspace defined by these two quantum numbers.
We need to determine
the parity for low-lying states with a given $S^z_T$.
We shall only be concerned
with chains of even length. For these chains
the ground-state is a singlet with even parity, $0^+$. Above the
ground-state is an exponentially low-lying triplet, $1^-$. In the
thermodynamic
limit the triplet and the singlet become degenerate and the
ground-state four-fold degenerate. This spectrum
can be seen to arise from the two
$S=1/2$ end-excitations
forming either an odd parity singlet or an even
parity triplet, in addition to an overall parity flip coming from the
rest of the ground-state. This
parity-flip can be understood
from the valence bond solid state~\cite{jphysc}
where we draw two valence bonds emanating
from each site. These valence
bonds represent singlet contractions of pairs of
$S=1/2$'s so they have a directionality associated with them. When we
make a parity transformation we flip the orientation of an odd number
of valence bonds resulting in a $(-)$ sign.
Thus, the parity, $P_E$, of a
state with no magnons present
is $(+)$ if the end excitations combine into
the singlet and $(-)$ for the triplet.
The parity of higher excited states,
containing one or more magnons, is a product of three factors,
$P_EP_{SW}P_m$.
$P_m$ contains a contribution of $(-)$ from each magnon
present.
This is because the magnons are created and annihilated by the
staggered magnetization operator, and this changes sign upon switching
even and odd sublattices. $P_{SW}$ is the parity of the spatial
wave-functions of the magnons. For instance,
for a single magnon,
the wave-functions, $\psi_i$, in Eq.~(\ref{eq:wf})
are $\psi_i=\sqrt{2/(L-1)}\sin k_ix$, $k_i=\pi n_i/(L-1)$, with
$n_i$ {\it odd} for even parity and
$n_i$ {\it even} for odd parity.
We take $0\leq x\leq L-1$ and parity
will therefore take $x$ into $L-1-x$.

For a chain with open boundary conditions the lowest lying state of a
given
magnetization, $M$, will have $M=m+1$, where $m$ is the number of
magnons present and the additional term, $1$, corresponds to the
end excitations forming a triplet.
In order to minimize the inter-magnon repulsion, the wave-function
for large $L$ takes the Bloch form of Eq.~(\ref{eq:wf}), with
$\psi_i$ as above and
$n_i=i$, in order to satisfy vanishing boundary conditions.
Thus Eq.~(\ref{eq:1}) becomes
\begin{equation}
E_{m+1}(L)-E_{1}(L)= (\Delta -h)m + {(v\pi )^2 \over 2\Delta
(L-1)^2}\sum_{i=1}^mn_i^2 + O(L^{-3}),\ \ n_i=i.
\label{eq:6}
\end{equation}
We test this formula below for $m=1$, $2$ and $3$.

The $1^-$ state becomes degenerate with the ground-state
in the thermodynamic limit
and we shall therefore
view this state as the reference
state and calculate energy gaps with respect to this state and not
$0^+$, as already implied in Eq.~(\ref{eq:6}).
We have calculated the gap as a function of chain length
between this state and three
of the low lying states using density matrices of the size
$243\times243$ keeping 81 eigenvectors of these matrices at each
iterations.
For each of these states
we have also calculated $<S^z_i>$ and $<{\bf S}_i\cdot{\bf S}_{i+1}>$
along a $L=100$ site chain using
a finite lattice method~\cite{white1}.
For a discussion of the numerical procedure we
refer the reader to Ref.~\onlinecite{white1}. A brief summary of our
results is shown in Table~\ref{spectrum}.

The lowest lying $M=m+1=2$ state corresponds to a state with the end
excitations in the $1^-$ state and one magnon present. This state
has therefore parity $(+)$ since $P_E=(-),P_{SW}=(+)$ and $P_m=(-)$.
We approximate the wave-function as consisting
of two factors existing
in different Hilbert spaces. A factor from the end
excitations and a factor, $\Psi_m$,
corresponding to the $m$-magnon wave-function.
{}From the above discussion we see that for a single magnon
the magnon part of the wave-function, $\Psi_m$, becomes
$
\Psi_1=\sqrt{\frac{2}{L-1}}\sin k_1x,\ \ n_1=1.
$
{}From Eq.~(\ref{eq:4}) we then see that
$
<S_i^z>_{2^+}-<S_i^z>_{1^-}=\frac{2}{L-1}\sin^2k_1 x,
$
which is shown as the solid line in Fig.~\ref{fig:1}.
Here the subtraction of $<S_i^z>_{1^-}$ essentially removes any
contribution from the end-excitations.
In Fig.~\ref{fig:1} we show $<S_i^z>_{2^+}-<S_i^z>_{1^-}$.
An excellent agreement is evident. Also
shown in Fig.~\ref{fig:1} is the local bond energy
$e^{21}_i=<{\bf S}_i\cdot{\bf S}_{i+1}>_{2^+}-
<{\bf S}_i\cdot{\bf S}_{i+1}>_{1^-}$.
The dilute boson model predicts an energy density, $e(x)\simeq \Delta
\sum_{j=1}^m\delta(x-\hat x_j)$
ignoring the $O(L^{-2})$ kinetic energy.
Therefore $e(x)$
should be proportional to $S^z(x)$ in Eq.~(\ref{eq:5}),
the proportionality
factor being the gap, $\Delta$. In Fig.~\ref{fig:1} this prediction is
shown as the dotted line.
{}From Eq.~(\ref{eq:6}) we
can now
extract values for the gap, $\Delta$, and the velocity, $v$. The fit
of $\Delta_{21}(L)=E_{2^+}(L)-E_{1^-}(L)$
to Eq.~(\ref{eq:6}) is excellent, and we obtain
$
\Delta_{21}(L)=0.4107(1)+74.7(4)(L-1)^{-2}+O([L-1]^{-3}),
$
with $\chi^2=4.55$. We see that $\Delta=0.4107(1)$, $v=2.49(1)$.
The value of $\Delta$ is in excellent agreement with what was
previously obtained~\cite{white1,white2}. The value of $v$
is in good agreement with the value
$v=2.46$ that can
be extracted from exact diagonalization~\cite{takahashi},
and the value $v\sim 2.36$ obtained from $1/S$
expansions~\cite{chubokov}. It is also in good agreement with the
experimental results on NENP~\cite{ma}, $v\sim2.45$.
The coefficient, $74.7(4)$, in front of the $(L-1)^{-2}$ term, that
determines $v$, differs marginally from what was obtained
by White~\cite{white1}, ($67.9$), due to the use of a different
polynomial form.

The lowest lying 2-magnon state has parity $(-)$ since $P_E=(-),
P_{SW}=(+), P_m=(+)$ and total magnetization M=m+1=3.
The magnon part of the wave-function is
$
\Psi_2=\frac{2}{L-1}[\sin k_1x_1\sin k_2x_2-\sin k_1x_2\sin k_2x_1]
\epsilon(x_1,x_2)$ with $ n_1=1,\ n_2=2.
$
Note that under parity $x_i\rightarrow
L-x_i-1, \sin k_1x_i$ is even, $\sin k_2x_i$ is odd and
$\epsilon(x_1,x_2)$ is odd resulting in $P_{SW}=(+)$.
We now obtain
$
<S_i^z>_{3^-}-<S_i^z>_{1^-}=
\frac{2}{L-1}\{\sin^2k_1x+\sin^2k_2x\},
$
which is shown as the solid line in Fig.~\ref{fig:2}.
The dotted line represents  the theoretical prediction for the local
bond
energy which is also shown in  Fig.~\ref{fig:2}.
Again we fit
$\Delta_{31}(L)=E_{3^-}(L)-E_{1^-}(L)$ to Eq.~(\ref{eq:6})
and we obtain
$\Delta_{31}(L)=0.823(1)+359(5)(L-1)^{-2}+O([L-1]^{-3})$. The constant
term should be $2\Delta$ in good agreement with the value of $\Delta$
obtained above.
Since in this case we expect $n_1=1,n_2=2$ and therefore
$\sum n_i^2=5$,
the coefficient in front of the $(L-1)^{-2}$ term should
be 5 times
greater than what we found for the $2^-$ level. Clearly this is the
case: we obtain $\sum n_i^2=4.80(6)$,
and the only values of $n_i$ consistent with our results are indeed
$n_1=1,n_2=2$.

The three magnon state with $M=m+1=4$ has parity $(+)$ by the same
arguments as above. The wave-function now has 6 terms and we obtain
$
<S_i^z>_{4^+}-<S_i^z>_{1^-}=
\frac{2}{L-1}\{\sin^2k_1x+\sin^2k_2x+sin^2k_3x\},
$
with $n_1=1,n_2=2,n_3=3$.
This expression is shown as the solid line in Fig.~\ref{fig:3},
along with the numerical results for
$<S_i^z>_{4^+}-<S_i^z>_{1^-}$ and the local bond energy
$<{\bf S}_i\cdot{\bf S}_{i+1}>_{4^+}-
<{\bf S}_i\cdot{\bf S}_{i+1}>_{1^-}$.
The dotted line is the prediction for the local bond energy.
Again good agreement is evident
between theory and the numerical results is seen. Fitting the
energy gap to Eq.~(\ref{eq:6}) we find that the $(L-1)^{-2}$ term now
has a coefficient of 1030(150). Thus in this case,
if we use the value of $v$ determined above,
$\sum n_i^2=14(2)$.
This is only consistent with the values $n_1=1, n_2=2, n_3=3$.
The DMRG method works progressively
worse for states with higher magnetization and
the agreement
between the theory and the numerical results is therefore
not as spectacular for this state as for the states previously
discussed.

In summary
we find that the numerical results are in excellent agreement
with Eq.~(\ref{eq:6})
with $v=2.49(1)$, $\Delta=0.4107(1)$. The magnons
behave as bosons with repulsive interactions among themselves and with
the end excitations. The lowest energy state for given $M$ has the
Bloch form of Eq.~(\ref{eq:wf}), implying the validity of
Eq.~(\ref{eq:ml}).

\acknowledgments
We would like
to thank Steven White for sendings us Refs. \onlinecite{white1}
and \onlinecite{white2} prior
to publication as well as for useful discussions.
This research was supported in part
by NSERC of Canada.

\begin{table}
\caption{The spectrum of the $L=100$
open $S=1$ antiferromagnetic Heisenberg
chain.}
\label{spectrum}
\begin{tabular}{cd}
\multicolumn{1}{c}{$S^P_T$}
&\multicolumn{1}{c}{$-E$}\\
\tableline
$1^-$& 138.940086 \\
$2^+$&138.522461\\
$3^-$&138.08557\\
$4^+$&137.603\\
\end{tabular} \end{table}

\begin{figure}
\caption{
The open circles represent
$<S_i^z>_{2^+}-<S_i^z>_{1^-}$. The solid line is the expression
given in the text.
Also shown, by triangles, is
$<{\bf S}_i\cdot{\bf S}_{i+1}>_{2^+}-
<{\bf S}_i\cdot{\bf S}_{i+1}>_{1^-}$.
The dotted line is the prediction
for this local bond energy.
}
\label{fig:1}
\end{figure}

\begin{figure}
\caption{
The open circles represent $<S_i^z>_{3^-}-<S_i^z>_{1^-}$.
The solid line is
the expression given in the text.
Also shown, by triangles, is
$<{\bf S}_i\cdot{\bf S}_{i+1}>_{3^-}-
<{\bf S}_i\cdot{\bf S}_{i+1}>_{1^-}$.
The dotted line is the prediction
for this local bond energy.
}
\label{fig:2}
\end{figure}

\begin{figure}
\caption{
The open circles represent $<S_i^z>_{4^+}-<S_i^z>_{1^-}$.
The solid line is
the expression given in the text.
Also shown, by triangles, is
$<{\bf S}_i\cdot{\bf S}_{i+1}>_{4^+}-
<{\bf S}_i\cdot{\bf S}_{i+1}>_{1^-}$.
The dotted line is the prediction
for this local bond energy.
}
\label{fig:3}
\end{figure}

\end{document}